# Phase diagrams and polarization reversal in nanosized $Hf_xZr_{1-x}O_{2-y}$

Eugene A. Eliseev[1], Yuri O. Zagorodniy[1], Victor N. Pavlikov[1], Oksana V. Leshchenko[1], Hanna V. Shevliakova[2], Miroslav V. Karpec[1], Andrei D. Yaremkevych[3], Olena M. Fesenko[3*], Sergei V. Kalinin[4†], and Anna N. Morozovska[3‡]

[1] Frantsevich Institute for Problems in Materials Science, National Academy of Sciences of Ukraine
Omeliana Pritsaka str., 3, Kyiv, 03142, Ukraine

2 - National Technical University of Ukraine "Igor Sikorsky Kyiv Polytechnic Institute",
pr. Beresteiskyi 37, 03056, Kyiv, Ukraine

[3] Institute of Physics, National Academy of Sciences of Ukraine,
pr. Nauky 46, 03028, Kyiv, Ukraine

[4] Department of Materials Science and Engineering, University of Tennessee,
Knoxville, TN, 37996, USA

**Abstract**

To describe the polar properties of the nanosized $Hf_xZr_{1-x}O_{2-y}$, we evolve the "effective" Landau-Ginzburg-Devonshire (LGD) model based on the parametrization of the Landau expansion coefficients for the polar and antipolar orderings. We have shown that the effective LGD model can predict the influence of screening conditions and size effects on phase diagrams, polarization reversal and structural properties of the nanosized $Hf_xZr_{1-x}O_{2-y}$ of various shape and sizes. To verify the model, we use available experimental results for $Hf_xZr_{1-x}O_2$ thin films and oxygen-deficient $HfO_{2-y}$ nanoparticles prepared at different annealing conditions. X-ray diffraction, which was used to determine the phase composition of the $HfO_{2-y}$ nanoparticles, revealed the formation of the ferroelectric orthorhombic phase in them. Micro-Raman spectroscopy was used to explore the correlation of lattice dynamics and structural changes appearing in dependence on the oxygen vacancies concentration in the $HfO_{2-y}$ nanoparticles. Since our approach allows to determine the conditions (shape, sizes, Zr content and/or oxygen vacancies amount) for which the nanosized $Hf_xZr_{1-x}O_{2-y}$ are ferroelectrics or antiferroelectrics, we hope that obtained results are useful for creation of next generation of Si-compatible ferroelectric gate oxide nanomaterials.

---

\* Corresponding author: fesenko.olena@gmail.com

† Corresponding author: sergei2@utk.edu

‡ Corresponding author: anna.n.morozovska@gmail.com

# I. Introduction

Ferroelectric memory elements like FeRAMs and FETs offer fast-switching and low-power consumption benefits, but face integration challenges with modern silicon-based CMOS technology [1, 2]. Lead-free binary hafnium ($HfO_2$) and zirconium ($ZrO_2$) oxides emerge as promising candidates for FeRAMs and FETs due to the discovery of ferroelectricity and antiferroelectricity in their thin films, which are Si-compatible. The binary oxide $Hf_{1-x}Zr_xO_2$ is recognized as next-generation Si-integrable materials according to the International Roadmap for Devices and Systems (IRDS™ 2021: Beyond CMOS) [1].

Bulk $HfO_2$ and $ZrO_2$ are high-k dielectrics without ferroelectric properties in a wide range of temperatures (below 1200 K) and pressures (below 12 GPa) that is confirmed by Raman spectroscopy [3, 4]. However, nanomaterials based on their solid solutions exhibit complex behavior influenced by various structural transitions. In particular, the ferroelectric properties observed in $Hf_{1-x}Zr_xO_2$ ($0 \leq x \leq 1$) thin films stem from the polar orthorhombic phase, which results in ferroelectric phase metastability disregarding its higher energy compared to the bulk nonpolar monoclinic phase. The properties of the $Hf_{1-x}Zr_xO_2$ thin films vary significantly depending on the factors like the substrate material, annealing conditions, deposition methods, film thickness, and dopants concentration [5, 6, 7]. Depending on these factors and x variation from 0 to 1, $Hf_{1-x}Zr_xO_2$ thin films can manifest as dielectric, ferroelectric, or antiferroelectric materials [8, 9].

Theoretical [10, 11, 12] and experimental [13, 14, 15] evidences underscore the crucial role of surface and grain boundary energies, as well as oxygen vacancies, for optimizing the $Hf_{1-x}Zr_xO_2$ nanomaterials for practical applications in advanced FeRAMs and FETs technologies. However, analytical research and optimization efforts are needed to fully exploit the potential of these binary oxides. In particular, the role of the Zr doping, oxygen vacancies, size, screening and surface effects in $Hf_xZr_{1-x}O_{2-y}$ nanoparticles are very poorly described theoretically, despite several experimental studies reveal their high potential for controllable synthesis [16, 17], nanoelectronics and capacitor technology [18], bio-safety and bio-medical applications [19, 20].

Here we consider the phase diagrams and polarization reversal in the nanosized $Hf_xZr_{1-x}O_{2-y}$ using the "effective" Landau-Ginzburg-Devonshire (**LGD**) model [21]. This approach is based on the parametrization of the Landau expansion coefficients for the polar (**FE**) and antipolar (**AFE**) orderings in $HfO_2$-based compounds from a limited number of polarization-field curves and hysteresis loops. To verify the model, we use available experimental results for $Hf_xZr_{1-x}O_2$ thin films to determine the Landau expansion coefficients. Using the coefficients, we calculate the polarization hysteresis and phase diagrams in $Hf_xZr_{1-x}O_2$ and oxygen-deficient $HfO_{2-y}$ nanoparticles, assuming that the oxygen vacancies can stabilize the polar orthorhombic phase. X-ray diffraction was used to determine the phase composition of the $HfO_{2-y}$ nanoparticles prepared at different annealing conditions. Micro-Raman spectroscopy was used to explore the correlation of lattice dynamics and structural changes appearing in the HfO$_{2-y}$ nanoparticles.

**II. The "effective" Landau-Ginzburg-Devonshire model for nanosized Hf$_x$Zr$_{1-x}$O$_{2-y}$**

To determine the spatial-temporal evolution of polarization in the nanosized Hf$_x$Zr$_{1-x}$O$_{2-y}$ we use the Kittel-type model [22] incorporating polar and antipolar modes [23, 24, 25] combined with the LGD approach [21]. Corresponding LGD free energy functional $F$ additively includes a bulk part – an expansion on the 2-th and 4-th powers of the polar ($P_f$) and antipolar ($A_f$) order parameters, $F_{bulk}$; a polarization gradient energy contribution, $F_{grad}$; an electrostatic contribution, $F_{el}$; and a surface energy, $F_S$. $F$ has the form [15, 21]:

$$F = F_{bulk} + F_{grad} + F_{el} + F_S, \quad (1)$$

where the constituent parts are

$$F_{bulk} = \int_{V_f} d^3r \left( \frac{a_P}{2} P_f^2 + \frac{b_P}{4} P_f^4 + \frac{\eta}{2} P_f^2 A_f^2 + \frac{a_A}{2} A_f^2 + \frac{b_A}{4} A_f^4 \right), \quad (2a)$$

$$F_{grad} = \int_{V_f} d^3r \frac{g_{ij}}{2} \left( \frac{\partial P_f}{\partial x_i} \frac{\partial P_f}{\partial x_j} + \frac{\partial A_f}{\partial x_i} \frac{\partial A_f}{\partial x_j} \right), \quad (2b)$$

$$F_{el} = - \int_{V_f} d^3r \left( P_i E_i + \frac{\varepsilon_0 \varepsilon_b}{2} E_i E_i \right), \quad (2c)$$

$$F_S = \frac{1}{2} \int_S d^2r \left( c_P P_f^2 + c_A A_f^2 \right). \quad (2d)$$

Here $V_f$ is the volume and $S$ is the surface of the nanosized Hf$_x$Zr$_{1-x}$O$_{2-y}$. Polarization vector is $\vec{P} = (P_1, P_2, P_f)$. The "effective" LGD expansion coefficients $a_P$, $b_P$, $a_A$, $b_A$ and $\eta$ are the functions dependent on Zr content "x" and oxygen deficiency "y"; $g_{ij}$ is the polarization gradient tensor, $c_P$ and $c_A$ are the surface energy coefficients; $\varepsilon_0$ is a universal dielectric constant, $\varepsilon_b$ is a background permittivity; $E_i$ are the electric field components ($i, j =$1, 2 and 3). For classical ferroelectric films with a pronounced temperature-dependent and strain-dependent soft mode, the coefficients $a_P$ and $a_A$ linearly depend on the temperature and strains (see e.g., Refs. [26, 27]). However, this is not the case for the Hf$_x$Zr$_{1-x}$O$_{2-y}$.

Spatial-temporal evolution of the polar and antipolar order parameters, $P_f$ and $A_f$, is determined from the coupled time-dependent LGD-type Euler-Lagrange equations, obtained by the minimization of the free energy $F$ [21]:

$$\Gamma_P \frac{\partial P_f}{\partial t} + a_P P_f + b_P P_f^3 + \eta A_f^2 P_f - g_{ij} \frac{\partial P_f}{\partial x_i \partial x_j} = E_3, \quad (3a)$$

$$\Gamma_A \frac{\partial A_f}{\partial t} + a_A A_f + b_A A_f^3 + \eta P_f^2 A_f - g_{ij} \frac{\partial A_f}{\partial x_i \partial x_j} = 0. \quad (3b)$$

$\Gamma_P$ and $\Gamma_A$ are Landau-Khalatnikov relaxation coefficients [28]. The relaxation times of $P_f$ and $A_f$ are $\tau_P = \Gamma_P/|a_P|$ and $\tau_A = \Gamma_A/|a_A|$, respectively.

The electric field component $E_3$ co-linear with the polarization $P_f$ is a superposition of external and depolarization fields, $E_3^e$ and $E_3^d$, respectively. The quasi-static field $E_3$ is related to the electric potential $\varphi$ as $E_3 = -\frac{\partial \varphi}{\partial x_3}$. The potential $\varphi$ satisfies the Poisson equation inside the nanosized $Hf_xZr_{1-x}O_{2-y}$. The potential $\varphi$ is fixed at the conducting electrodes, obeys the Debye-Hukkel equation in the semiconducting electrodes covering the film, or inside the screening shell covering the nanoparticle.

For the single-domain film the analytical expressions for the electric field components are available:

$$E_3^d = \frac{-\lambda}{\varepsilon_0(\varepsilon_s h + \varepsilon_b \lambda)} P_f, \quad E_3^e = \frac{\varepsilon_s}{\varepsilon_s h + \varepsilon_b \lambda} U, \qquad \text{(film)} \qquad (4a)$$

where $h$ is the thickness of $Hf_xZr_{1-x}O_{2-y}$ film, $\lambda$ is the effective Debye-Hukkel screening length in the semiconducting electrodes, which is usually very small (much less than 1 nm), $\varepsilon_s$ is the lattice dielectric permittivity of the electrodes, $U$ is the voltage applied between the electrodes (see **Fig. 1(a)**).

For the single-domain spherical nanoparticle the analytical expressions for the electric field components have the form:

$$E_3^d = -\frac{1}{\varepsilon_b + 2\varepsilon_s + \varepsilon_s(R/\lambda)} \frac{P_f}{\varepsilon_0}, \quad E_3^e = \frac{3\varepsilon_s}{\varepsilon_b + 2\varepsilon_s + \varepsilon_s(R/\lambda)} E_3^0, \qquad \text{(nanoparticle)} \qquad (4b)$$

where $R$ is the radius of the $Hf_xZr_{1-x}O_{2-y}$ nanoparticle, $\lambda$ is the effective screening length in the screening shell with the relative dielectric permittivity $\varepsilon_s$ (see **Fig. 1(b)**). Here, $\lambda$ also can be rather small (less than 0.1 – 1 nm) due to free charges and surface band bending in the shell. Only if $\lambda \gg R$ and $\varepsilon_s \sim \varepsilon_b$, the field in the nanoparticle core is of the same order as the applied field $E_3^0$. The derivation of Eqs.(4) is given in Ref. [29].

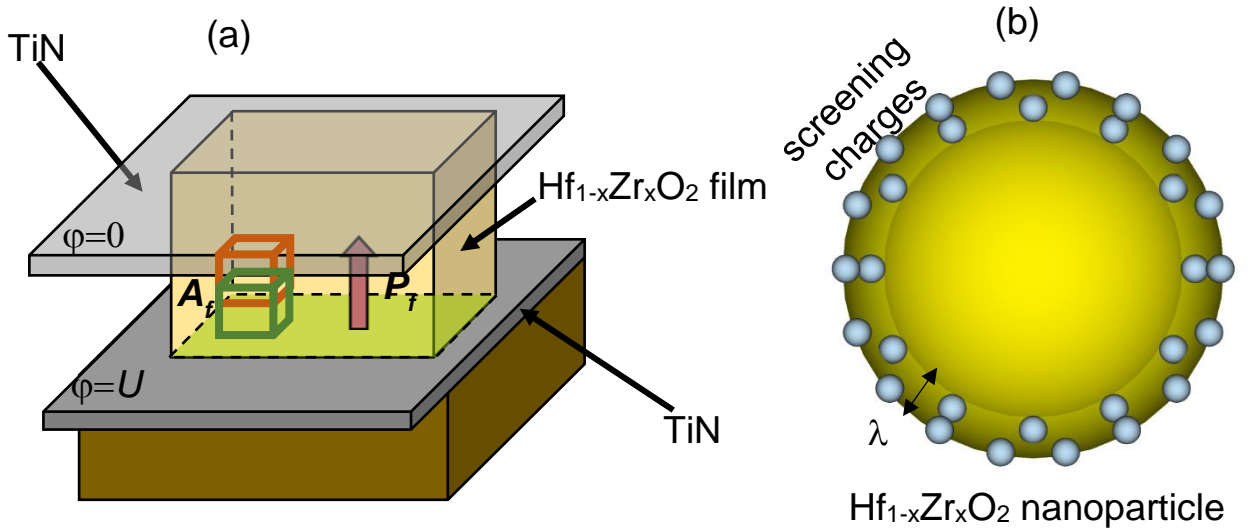

**Figure 1**. **(a)** The geometry of the considered heterostructure, consisting of a $Hf_xZr_{1-x}O_2$ film of thickness $h$ placed between conductive TiN electrodes. **(b)** The radial cross-section of the $Hf_xZr_{1-x}O_2$ nanoparticle covered with the shell of screening charge with the effective screening length $\lambda$.

Corresponding boundary conditions to Eqs.(3) are of the third kind [30]:

$$\left(c_P P_f - g_{ij} n_i \frac{\partial P_f}{\partial x_j}\right)\bigg|_S = 0, \quad \left(c_A A_f - g_{ij} n_i \frac{\partial A_f}{\partial x_j}\right)\bigg|_S = 0. \tag{5}$$

Here $\vec{n}$ is the outer normal to the $Hf_xZr_{1-x}O_{2-y}$ surface S. For the model case of a diagonal polarization gradient tensor, $g_{ij} = g\delta_{ij}$, it is convenient to introduce the extrapolation lengths, $\Lambda_P = \frac{g}{c_P}$ and $\Lambda_A = \frac{g}{c_A}$ are, which physical range is (0.5 – 5) nm [31].

To apply the analytical expressions (4) we should assume that $\Lambda_{P,A} \to \infty$ and λ is very small (much less than 1 nm), and thus the free charges in the electrodes (or in the shell) provide an effective screening of the $Hf_xZr_{1-x}O_{2-y}$ spontaneous polarization and prevent the domain formation, so that the assumption of the single-domain state in $Hf_xZr_{1-x}O_{2-y}$ is self-consistent. For higher λ one should use the finite element modeling (**FEM**) to account for the possible domain formation.

As a rule, the polar order parameter is observable (i.e., measurable), and the antipolar order parameter cannot be directly measured. However, the nonlinear coupling between $P_f$ and $A_f$ changes the field dependence $P_f(E_3)$.

### A. Thin films of $Hf_xZr_{1-x}O_2$

To verify the effective LGD model, we use available experimental results for $Hf_xZr_{1-x}O_2$ thin films at room temperature [8]. The geometry of the considered heterostructure, consisting of a $Hf_xZr_{1-x}O_2$ film of thickness $h$ placed between conductive TiN electrodes is shown in **Fig. 1(a)**. The temperature dependence of the polar properties of Zr-doped $HfO_2$ is considered elsewhere [32].

Examples of how the LGD model works quantitatively are shown in **Figs. A1** and **A2** in Supplement [33]. Here polarization hysteresis loops, measured experimentally in $Hf_xZr_{1-x}O_2$ thin films by Park et al. [8], are shown. The films thickness was 9.2 nm and the Zr content "x" varied from 100 % to 50 %. The films were covered with conducting TiN electrodes. Fits of dielectric, paraelectric, antiferroelectric, and ferroelectric loops show that the proposed effective LGD free energy qualitatively and semi-quantitatively describes the experimental results for the case of perfect screening (λ = 0). The LGD-model parameters, determined from fitting of experimental results from Park et al. [8], are listed in the last four columns of **Table AI** in **Appendix A** [33]**.**

The composition dependences $a_P(x)$, $b_P(x)$, $b_A(x)$, $a_A(x)$, and $\eta(x)$, determined from the fitting of experimental results [8] and further interpolated entire the range $0 \leq x \leq 1$, are shown in **Fig. 2(a)** and **2(b)**. To interpolate the points in the figures, polynomial x-functions are used for the dependence of LGD expansion coefficients on Zr content $x$:

$$a_P(x) = 3.22663 \cdot 10^9 + 2.66096 \cdot 10^{10}x - 5.0692 \cdot 10^{11}x^2 + 2.3960 \cdot 10^{12}x^3 - 5.0558 \cdot 10^{12}x^4 + 4.9335 \cdot 10^{12}x^5 - 1.7907 \cdot 10^{12}x^6, \tag{6}$$

$$b_P(x) = b_A(x) = 1.4535 \cdot 10^{10} - 2.23815 \cdot 10^{11}x + 1.7468 \cdot 10^{12}x^2 - 4.6296 \cdot 10^{12}x^3 + 4.0586 \cdot 10^{12}x^4, \tag{7}$$

$$a_A(x) = 1.0924 \cdot 10^9 - 1.5215 \cdot 10^{10}x + 2.5362 \cdot 10^{10}x^2, \tag{8}$$

$$\eta(x) = 6.0712 \cdot 10^9 + 7.5685 \cdot 10^{10}x - 4.6661 \cdot 10^{10}x^2. \tag{9}$$

Notably that Eqs.(6)-(9) are valid in the narrow range of the $Hf_xZr_{1-x}O_2$ sizes, e.g., for the film thickness range 5 nm $\leq h \leq$ 15 nm, because thinner films lose their ferroelectric properties and thicker films are described by another set of LGD coefficients (see other experimental results in Ref.[8] and their fitting in Ref.[21]).

Using the free energy (1), expressions for the electric field (4a) and x-dependences of the LGD coefficients (6)-(9) we calculated the phase diagram of the $Hf_xZr_{1-x}O_2$ thin films in dependence on the Zr content $x$ and the ratio $h/\lambda$ (see **Fig. 2(c)**). It is seen from the diagram that the increase of $x$ from 0 to 0.17 leads to the transition of the dielectric (DE) state to the antiferroelectric (AFE) phase, then to the mixed ferrielectric (FEI) state for x ≈ 0.5, and then to the ferroelectric (FE) phase for $x$>0.5. The further increase of $x$ from 0.6 to 0.7 leads to the gradual disappearance of ferroelectricity, and to the appearance of the paraelectric (PE) phase at $x$>0.7, which continuously transforms to the reentrant DE state for $x$>0.8. At the same time the diagram become $h$-dependent for $h/\lambda > 10$, since the contribution of the depolarization field becomes negligibly small with the increase of the ratio $h/\lambda$ (see Eq.(4a)). The condition of "weak" screening, $5 < h/\lambda < 10$, is the actual range of the film thickness effect manifestation, as well as the domain formation is possible exactly in the range of $h/\lambda$.

It may seem that the diagram in **Fig. 2(c)** is applicable for arbitrary film thickness, and the phase boundaries depend on the ratio $h/\lambda$ and Zr fraction $x$. However, we need to remind readers that the $x$-dependence of LGD coefficients given by Eqs.(6)-(9) are valid for 5 nm $\leq h \leq$ 15 nm. Thus, **Fig. 2(c)** is the diagram showing the strong influence of the Zr content and weaker influence of the screening effects on the phase state of $Hf_xZr_{1-x}O_2$ thin films.

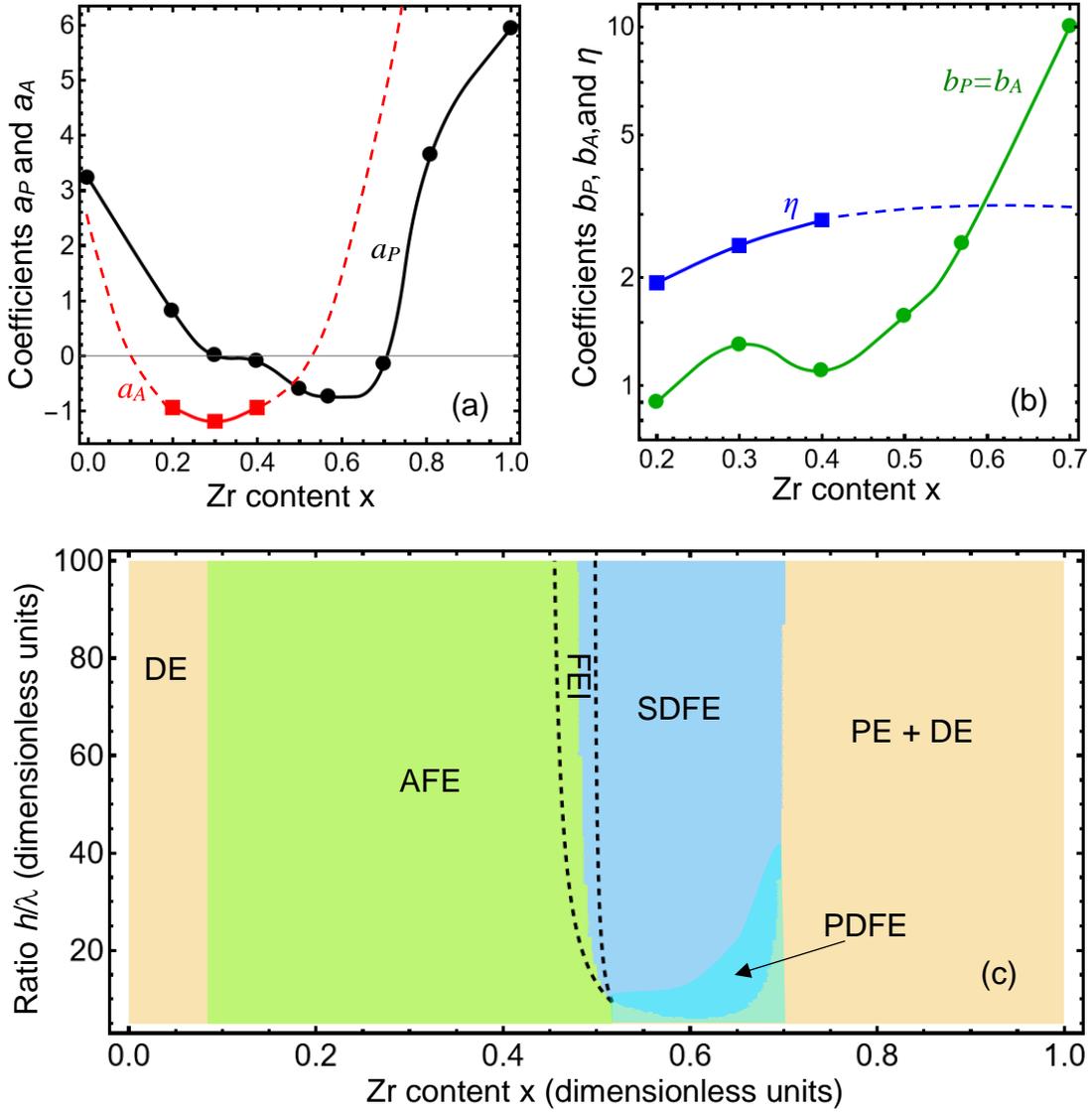

**Figure 2**. The composition dependence of the effective LDG coefficients $a_P$ and $a_P$ (in $10^9$ F/m) **(a)**, as well as $b_P = b_A$ and $\eta$ (in $10^{10}$ Vm$^5$/C$^3$) **(b)** determined from the fitting of hysteresis loops of Hf$_x$Zr$_{1-x}$O$_2$ thin films, shown in **Figs. A1-A2**. **(c)** Phase diagram of the Hf$_x$Zr$_{1-x}$O$_2$ film calculated in dependence of the Zr content x and the ratio $h/\lambda$. Parameters $\varepsilon_s = 10$ and $\varepsilon_b = 3$. The abbreviations "DE", "AFE", "FEI", "SDFE" and "PE" refer to the dielectric, paraelectric, antiferroelectric, mixed ferrielectric, and single-domain ferroelectric phases, respectively. A possible region of the poly-domain ferroelectric (PDFE) state stability is located inside the semi-transparent cyan area.

The effective LGD model with determined coefficients (6)-(9) can predict the phase diagrams, ground and metastable states, polar and structural properties of the nanosized Hf$_x$Zr$_{1-x}$O$_{2-y}$ of various shape and sizes from 5 to 50 lattice constants. Below we apply the effective LGD model to the spherical Hf$_x$Zr$_{1-x}$O$_2$ nanoparticles covered with a screening shell.

## B. Nanoparticles of $Hf_xZr_{1-x}O_2$

Polarization-field dependences of the $Hf_xZr_{1-x}O_2$ nanoparticles calculated for different Zr content x are shown in **Fig. 3**. The particles have the spherical shape, 5-nm radius and are well-screened ($\lambda \ll 0.1$ nm). Black dotted curves show the quasi-equilibrium field dependences of the antipolar order parameter, $A_f(E)$, and colored solid curves show the polarization-field dependences, $P_f(E)$. It is seen from the figure that the increase of Zr content x from 0 to 0.6 leads to the gradual transitions of the $P_f(E)$ loop shape from the dielectric-type to the antiferroelectric-type, then to the mixed ferrielectric-type and then to ferroelectric-type loops. The further increase of x from 0.6 to unity leads to the ferroelectricity disappearance (at x>0.7) and to the appearance of the paraelectric-type and dielectric-type polarization response to the external field. Notably that the sequence "dielectric-antiferroelectric-ferrielectric-ferroelectric-paraelectric-dielectric" response of $P_f(E)$ was observed experimentally [8] for the 9.2-nm $Hf_xZr_{1-x}O_2$ films covered by conducting electrodes. Since we used the LGD coefficients, which describe the experimental trend [8] for the well-screened 9.2-nm $Hf_xZr_{1-x}O_2$ films and the chosen nanoparticle size (10 nm) is very close to the 9-nm film thickness, results shown in **Fig. 3** are consistent.

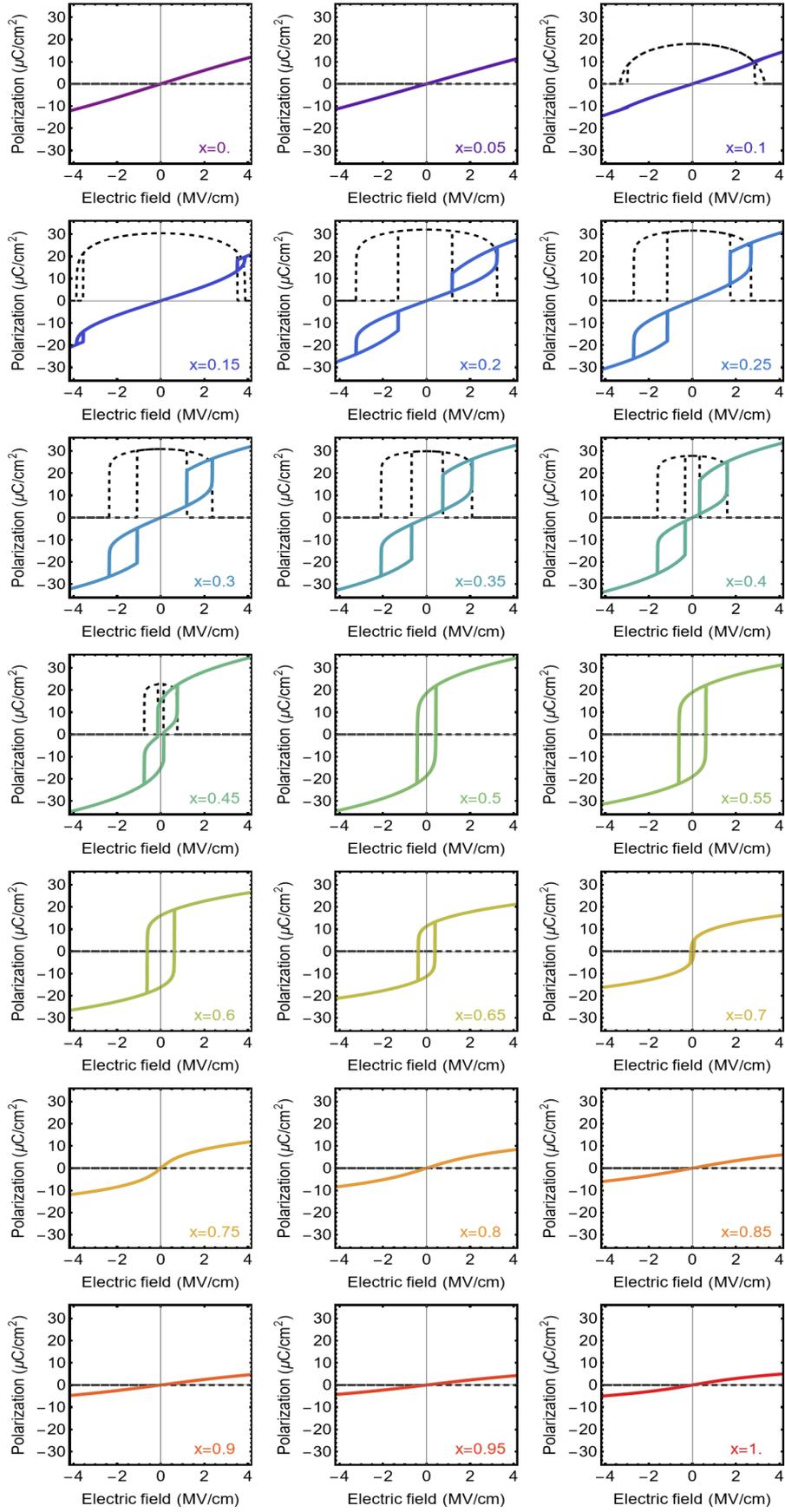

**Figure 3**. The quasi-equilibrium dependences of polar ($P_f$) and antipolar ($A_f$) order parameters on the external field $E$ calculated for the Hf$_x$Zr$_{1-x}$O$_2$ nanoparticles, which Zr content $x$ changes from 0 to 1 with the step 0.05

(see legends in the plots). Parameters $\varepsilon_s = 10$, $\varepsilon_b = 3$, $\lambda = 0.01$ nm, and $R = 5$ nm. The dielectric-type, paraelectric-type, antiferroelectric-type and ferroelectric-type loops are shown. Color curves are $P_f$ and black dotted curves are $A_f$.

To study the screening and size effects, the ratio $R/\lambda$ should be changed (see e.g., Eq.(4b)). Phase diagrams of the spherical $Hf_xZr_{1-x}O_2$ nanoparticles, calculated in dependence of Zr content x and the ratio $R/\lambda$ for several values of the shell permittivity $\varepsilon_s = 1$ (air), $\varepsilon_s = \varepsilon_b = 3$ (core and shell have the same dielectric constant), $\varepsilon_s = 10$ (the high-k dielectric shell) and $\varepsilon_s = 30$ (the paraelectric shell), are shown in **Fig. 4(a)**, **4(b)**, **4(c)** and **4(d)**, respectively.

It is seen from the diagram for $\varepsilon_s = 1$ that the increase of Zr content $x$ from 0 to 0.1 leads to the transition from the DE state to the AFE phase (see **Fig. 4(a)**). The further increase of $x$ from 0.1 to 0.5 leads to the gradual degradation (for $x>0.3$) and eventual disappearance of antiferroelectricity (at $x>0.52$), and to the appearance of the reentrant DE state.

It is seen from the diagrams, shown in **Figs. 4(b)-4(d)**, that the increase of $\varepsilon_s$ from 3 to 30 leads to the appearance of the FEI state and the FE phase regions, which areas increase with $\varepsilon_s$ increase. Independently on $\varepsilon_s$, the increase of Zr content $x$ from 0 to 0.1 leads to the transition of the DE state to the AFE phase. The increase of $x$ above 0.5 leads to the appearance of the FEI state for $R/\lambda$ more than a critical value (see a thin region located between the black dashed curves), or to the PDFE state appearance for $R/\lambda$ less than the critical value. The critical value decreases from $R/\lambda \approx 40$ (see **Fig. 4(b)** for $\varepsilon_s = 3$) to 10 (see **Fig. 4(c)** for $\varepsilon_s = 10$), and then disappears for further increase of $\varepsilon_s$ (see **Fig. 4(d)** for $\varepsilon_s = 30$). The FEI phase transforms to the FE phase with $x$ increase. The further increase of $x$ from 0.6 to 1 leads to the ferroelectricity disappearance at $x>0.7$, and to the appearance of the PE phase, which continuously transforms to the DE state for $x>0.8$.

All diagrams in **Fig. 4** become $R$-independent for $R/\lambda \gg 10$, since the contribution of the depolarization field becomes negligibly small with the $R/\lambda$ increase (see Eq.(4b)). The range of the incomplete screening conditions, $5 < R/\lambda < 50$, is the actual range of the size effect existence. Notably, the domain formation is possible for the weak and incomplete screening (see the semi-transparent cyan area of PDFE states in **Figs. 4(b)** and **4(c)**).

It may seem that the diagrams are applicable for arbitrary radius $R$ and the phase boundaries depend on the $R/\lambda$ and $x$. However, the x-dependence of LGD coefficients given by Eqs.(6)-(9) is valid in the narrow range of sizes, i.e., for 2.5 nm $\leq R \leq$ 7.5 nm. Thus, one should consider **Figs. 4** as the diagrams showing the influence of the Zr content and screening effects on the phase state of the $Hf_xZr_{1-x}O_2$ nanoparticles.

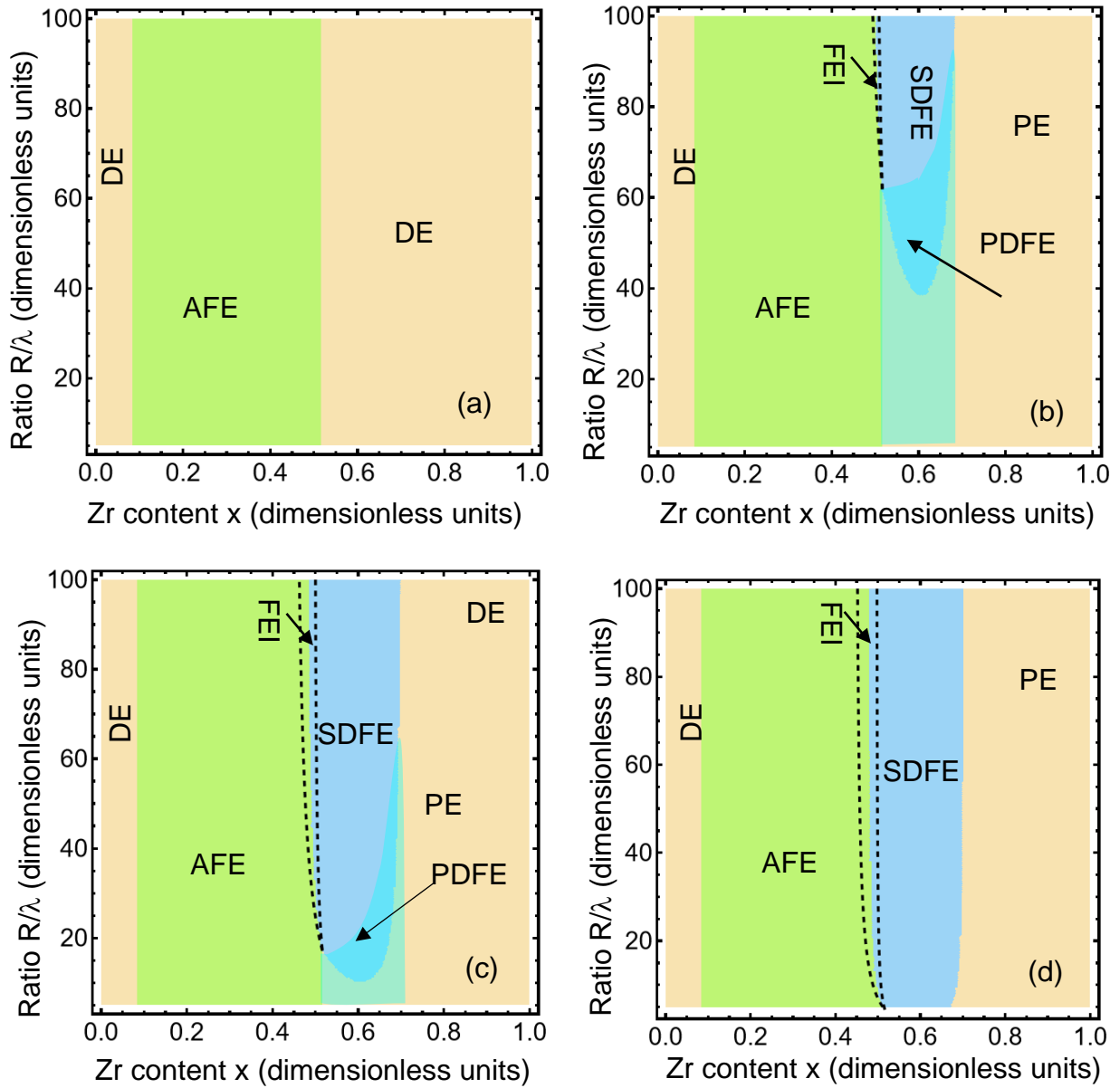

**Figure 4**. Phase diagrams of the spherical $Hf_xZr_{1-x}O_2$ nanoparticles calculated in dependence of Zr content x and the ratio $R/\lambda$. The abbreviations "DE", "AFE", "FEI", "SDFE" and "PE" refer to the dielectric, paraelectric, antiferroelectric, mixed ferrielectric, and single-domain ferroelectric phases, respectively. A possible region of the poly-domain ferroelectric (PDFE) state stability is located inside the semi-transparent cyan area. Parameters $\varepsilon_s = 1$ **(a)**, $\varepsilon_s = 3$ **(b)**, $\varepsilon_s = 10$ **(c)**, and $\varepsilon_s = 30$ **(d)**, and $\varepsilon_b = 3$.

### III. X-ray diffraction and Raman spectra of $HfO_{2-y}$ nanoparticles

It is known that oxygen vacancies in oxides cause the effect of the so-called "chemical pressure" (another name is the Vegard stresses) of the crystal lattice [34]. The stresses affect the formation of structural states unstable under normal conditions in $HfO_2$ and $ZrO_2$ oxides, which have from 3 to 4 polymorphic forms. Thus, the Vegard stresses induced by the oxygen vacancies can shift the stability conditions and lead to the metastability of the orthorhombic polar phase in oxygen-deficient $HfO_{2-y}$ nanoparticles. To verify the idea, we prepared several samples of stoichiometric $HfO_2$ nanoparticles and oxygen-deficient $HfO_{2-y}$ nanoparticles, which differ by the annealing conditions. The nanopowder samples were obtained in two ways of synthesis.

(1) By organonitrate synthesis from mixtures of as-prepared Hf hydroxide, ammonium nitrate and dextrin. The heating temperature of the mixtures was maintained in the range of (400 – 700)°C. To create reducing conditions for the synthesis, an excess of the organic additive relative to the $NO_2$ oxidant was used. Below we discuss results of experimental studies for the group of white-colored samples, named as the "**Sample 1**", which were prepared by the way (1), namely annealed at 700°C for 6 hours in air.

(2) By pyrogenic synthesis from the same mixtures of hydroxides washed from nitrates with the addition of dextrin in the $CO+CO_2$ environment heated in the temperature range of (400 – 700)°C for 6-20 hours. Excess of carbon from the powders obtained in this way was removed by the short-term exposure (10-15 minutes) in air at the temperature (450 – 500)°C. Below we discuss results of experimental studies for the 3 groups of grey-colored samples, named as the "**Sample 1, 2, 3**", which were prepared by the way (2). Note that the **Sample 2**, annealed at 700°C for 6 hours in $CO+CO_2$ atmosphere, has a light grey color; the **Sample 3**, annealed at 650°C for 16 hours in $CO_2$ atmosphere, has a grey color; and the **Sample 4**, annealed at 600°C for 16 hours in $CO+CO_2$ atmosphere, has a dark-grey color.

X-ray diffraction (XRD) was used to determine the phase composition of the $HfO_{2-y}$ nanoparticles. We use an XRD-6000 diffractometer with Cu-K$\alpha$1 radiation, and the measured angle (2$\theta$) was from 5° to 70°. To identify the crystallographic phases in the studied system we used the database of the International Committee for Powder Diffraction Standards (JCPDS PDF-2).

Micro-Raman spectroscopy was used to explore the correlation of lattice dynamics and structural changes appearing in dependence on the oxygen vacancies concentration. Raman spectra were measured using a Renishaw InVia (England) micro-Raman spectrometer equipped with a DM2500 Leica confocal optical microscope. A laser operating at a wavelength of $\lambda = 633$ nm was used to measure the Raman scattering spectra. Processing of Raman spectra was performed using the WiRE 5.2 program, which was used to determine the peaks and decompose the bands into components. All measurements were performed at room temperature.

The XRD spectra of the Sample 1, shown in **Fig.5(a)**, reveals the pure monoclinic phase ("m"), as anticipated for a stoichiometric $HfO_2$ nanopowder. The average size of coherent scattering regions with the monoclinic symmetry is 13 nm. The Raman spectra of the Sample 1, shown in **Fig.5(b)**, has many sharp peaks (not less than 8) located below the 800 cm$^{-1}$, and then the Raman signal intensity strongly increases and saturates above the 800 cm$^{-1}$. The saturation can be related with the significant amount of Raman-active luminescence centers.

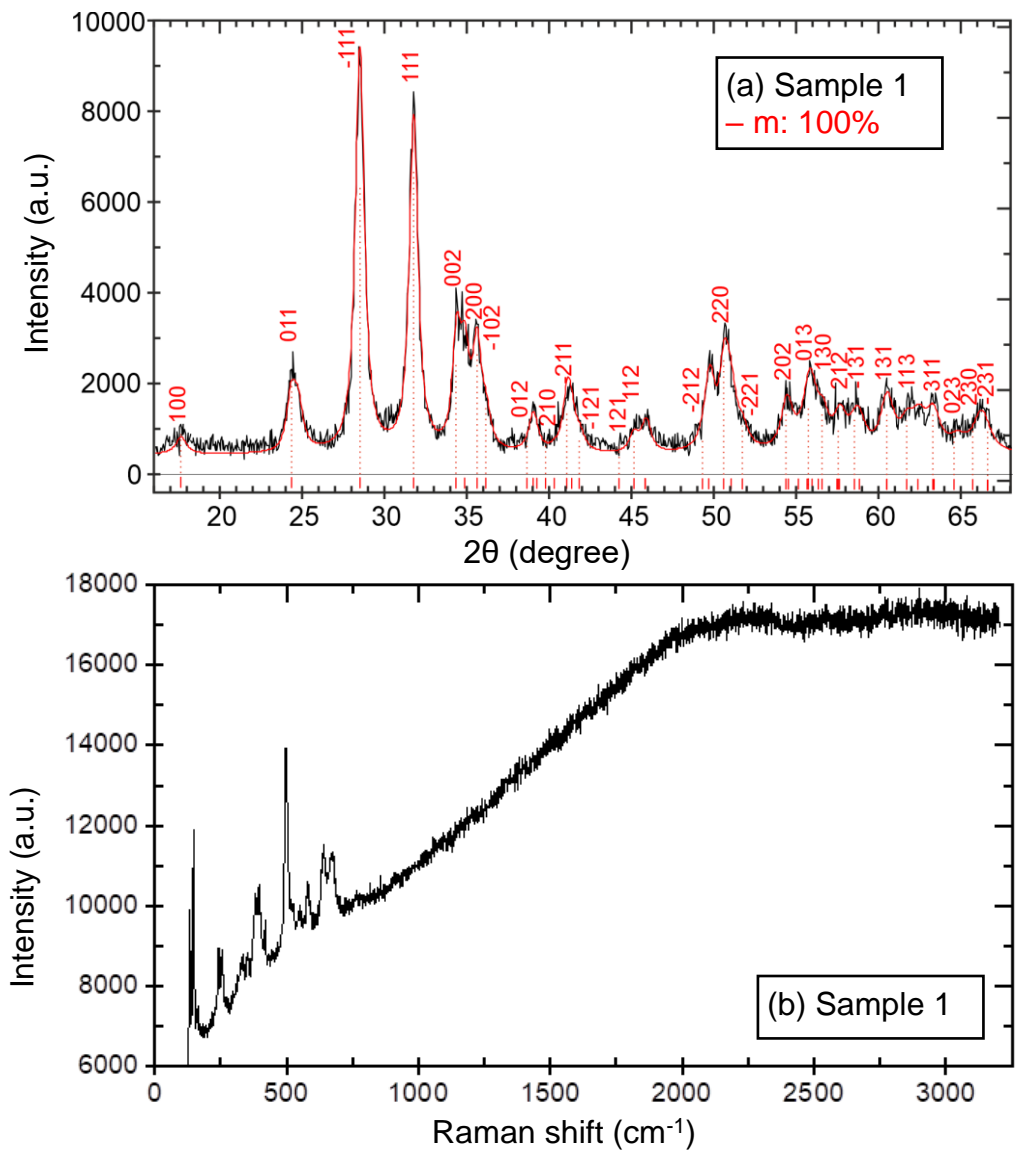

**Figure 5**. XRD **(a)** and Raman **(b)** spectra of the nanopowder Sample 1 consisting of stoichiometric $HfO_2$ nanoparticles.

The XRD spectra, shown in **Fig.6(a)-(c)**, correspond to the oxygen-deficient $HfO_{2-y}$ Samples 2, 3 and 4, respectively. The XRD spectra demonstrate the gradual increase of the orthorhombic phase ("o61"), and the gradual decrease of the monoclinic ("m") occurring under the change of annealing conditions. In particular, the fraction of non-FE monoclinic phase gradually decreases from 68.2 % to 27.3%, and the fraction of FE orthorhombic phases gradually increases from 32.0% to 72.7% for the Samples 2, 3 and 4, respectively. We relate the change in the phase composition with the gradual increase of the oxygen vacancies concentration, which appear due to the change of the annealing conditions. Simultaneously with the composition change the average size of coherent scattering regions, which have the orthorhombic symmetry, varies from 10 nm to 14 nm, and the average size of coherent scattering regions, which have the monoclinic symmetry, varies from 9 nm to 23 nm (see **Table A3** in **Appendix A** [33]).

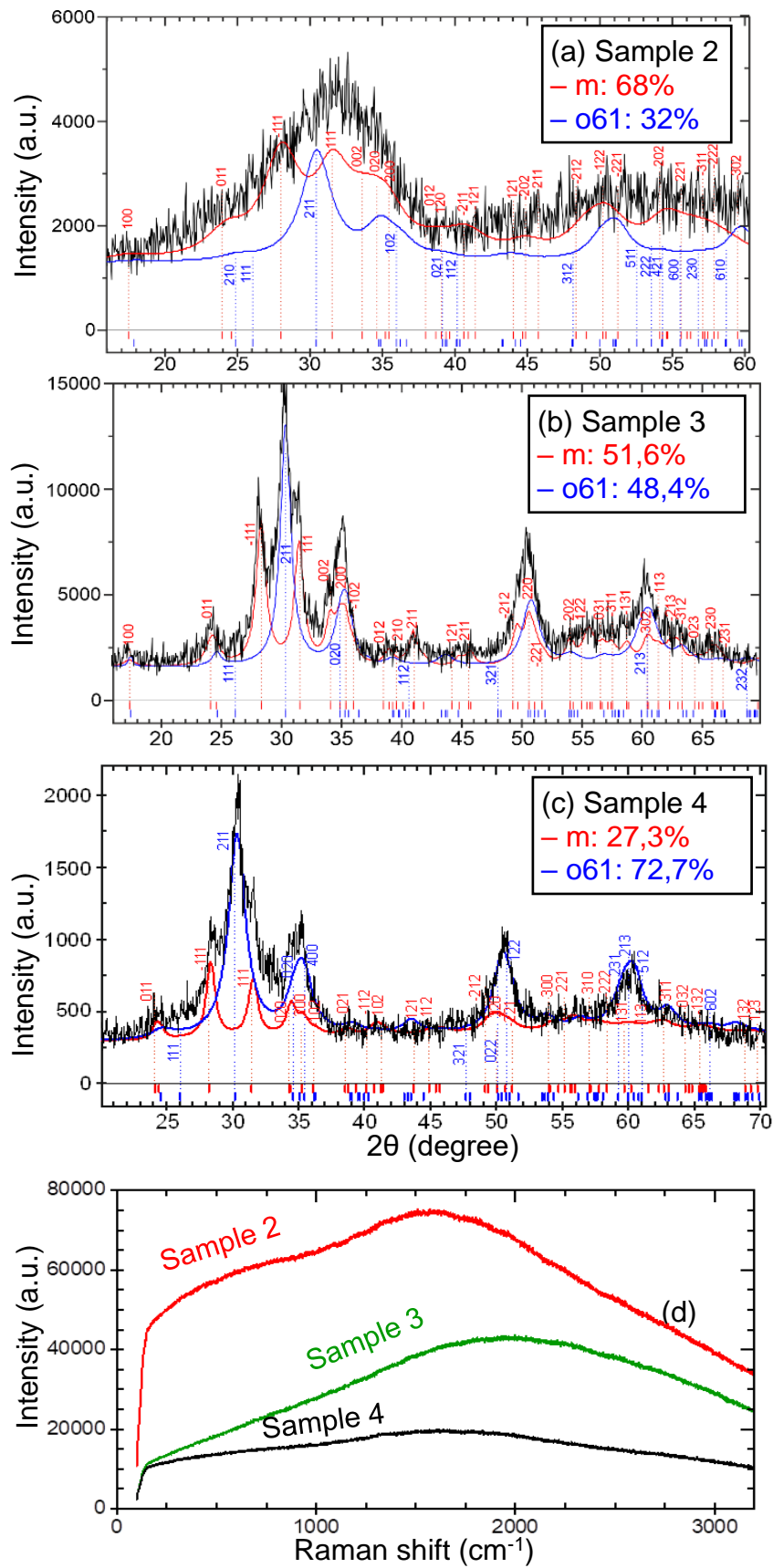

**Figure 6**. XRD **(a, b, c)** and Raman **(d)** spectra of nanopowder Samples 2-4 consisting of the oxygen-deficient $HfO_{2-y}$ nanoparticles.

The Raman spectra of the Samples 2-4, shown in **Fig.6(d)**, do not have any sharp peaks located below 3500 cm$^{-1}$. Instead, the Raman signal intensity reveal a very diffuse maxima located between (1000 – 2500) cm$^{-1}$. The maxima height gradually decreases from the Sample 2 to the Samples 3 and 4, respectively. The behavior is characteristic for the very high amount of Raman-active luminescence centers, which concentration increases due to the increase of oxygen vacancies concentration in the samples.

Hence, the X-ray diffraction, complemented by the Raman spectroscopy, revealed the formation of the ferroelectric orthorhombic phase under the increase of oxygen vacancies amount in the HfO$_{2-y}$ nanoparticles prepared at different annealing conditions. The determined fractions of the monoclinic and orthorhombic phases along with the sizes of the coherent scattering regions allow us to use the information for the application of effective LGD model. Obtained results are discussed in the next section.

### IV. Polar properties of the oxygen-deficient HfO$_{2-y}$ nanoparticles

Using the fractions of the monoclinic and orthorhombic phases along with the sizes of the coherent scattering regions, which were determined from the XRD data, we apply the effective LGD model to the stoichiometric and oxygen-deficient HfO$_{2-y}$ nanoparticles. We regard that the concentration of oxygen vacancies is maximal at the surface of quasi-spherical nanoparticle and monotonically decreases towards its center (see **Fig. 7(a)**). The vacancies, being elastic dipoles, create elastic Vegard strains [35, 36], which increase the stability of the FE orthorhombic phase due to electrostriction coupling [11, 15, 21].

The phase diagram of the spherical HfO$_{2-y}$ nanoparticles, calculated in dependence of oxygen vacancies amount y and the ratio $R/\lambda$ for $\varepsilon_s = 5$ is shown in **Fig. 7(b)**. It is seen from the diagram, that the increase of $y$ from 3 % to 4.8 % leads to the transition of the AFE state to the single-domain (SDFE) or poly-domain (PDFE) ferroelectric phase. The thin region of FEI states separates the AFE state from the SDFE phase for $R/\lambda$ more than the critical value, $R/\lambda \approx 40$ (see the thin region located between the black dashed curves). For $R/\lambda$ less than the critical value, the AFE state transforms into the PDFE with $y$ increase. The further increase of $y$ above 4.8 % leads to the ferroelectricity disappearance, and to the appearance of the PE phase, which continuously transforms to the DE state for further increase of y above 7 % (not shown in **Fig. 7(b)**). The appearance of the FE phase in the oxygen-deficient HfO$_{2-y}$ nanoparticles agrees with the experimental XRD results presented in the previous section.

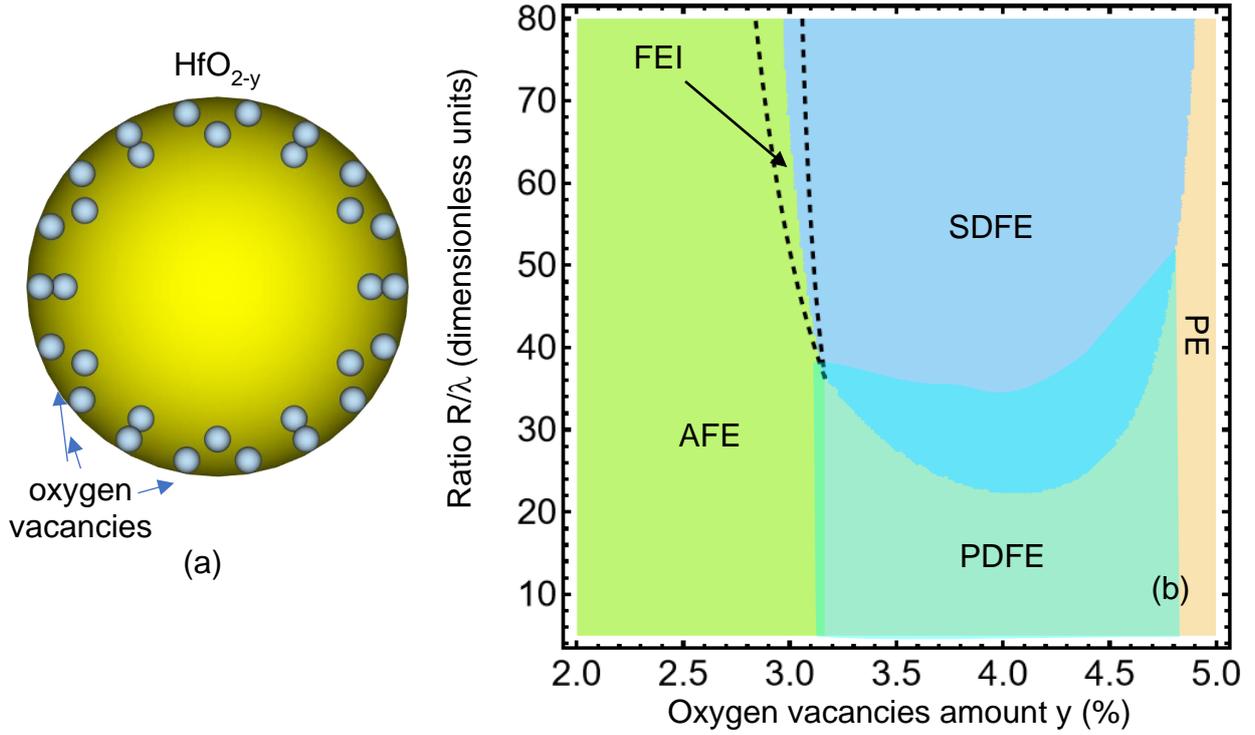

**Figure 7.** (a) The radial cross-section of the HfO$_{2-y}$ nanoparticle. (b) Phase diagrams of spherical HfO$_{2-y}$ nanoparticles calculated in dependence of oxygen vacancies amount y and the ratio $R/\lambda$. The abbreviations "AFE", "FEI", "SDFE" and "PE" refer to the paraelectric, antiferroelectric, mixed ferrielectric, and single-domain ferroelectric phases, respectively. A possible region of the poly-domain ferroelectric (PDFE) state stability is located inside the semi-transparent cyan area; $\varepsilon_s = 5$ and $\varepsilon_b = 3$.

## V. Conclusions

We have shown that the effective LGD model can predict the influence of screening conditions and size effects on phase diagrams, polarization reversal and structural properties of the nanosized Hf$_x$Zr$_{1-x}$O$_{2-y}$ of various shape and sizes. The effective LGD parameters was determined from the available experimental results for Hf$_x$Zr$_{1-x}$O$_2$ thin films.

We prepare oxygen-deficient HfO$_{2-y}$ nanoparticles, where the XRD revealed the formation of the ferroelectric orthorhombic phase. Micro-Raman spectroscopy was used to explore the correlation of lattice dynamics and structural changes appearing in the nanoparticles. We apply the effective LGD model to the HfO$_{2-y}$ nanoparticles and explain the appearance of the ferroelectric orthorhombic phase in dependence on the oxygen vacancies concentration.

Since our approach allows to determine the conditions (shape, sizes, Zr content and/or oxygen vacancies amount) for which the nanosized Hf$_x$Zr$_{1-x}$O$_{2-y}$ are ferroelectrics or antiferroelectrics, we hope that obtained results are useful for creation of next generation of Si-compatible ferroelectric gate oxide nanomaterials.

**Data availability statement.** Numerical results presented in the work are obtained and visualized using a specialized software, Mathematica 14.0 [37].

**Authors' contribution.** The research idea belongs to A.N.M., Y.O.Z. and S.V.K. E.A.E. wrote the codes, performed numerical calculations and compare with experiments jointly with H.V.S. V.N.P., O.V.L. and Y.O.Z. prepared the samples. M.V.K. performed XRD measurements. O.M.F. and A.D.Y. performed Raman measurements. A.N.M. formulated the theoretical problem, performed analytical calculations, and wrote the manuscript draft. All co-authors analyzed the obtained results and corresponding authors worked on the manuscript improvement.

**Acknowledgments.** The work (A.N.M.) is supported by the DOE Software Project on "Computational Mesoscale Science and Open Software for Quantum Materials", under Award Number DE-SC0020145 as part of the Computational Materials Sciences Program of US Department of Energy, Office of Science, Basic Energy Sciences. This effort (S.V.K) was primarily supported by the center for 3D Ferroelectric Microelectronics (3DFeM), an Energy Frontier Research Center funded by the U.S. Department of Energy (DOE), Office of Science, Basic Energy Sciences under Award Number DE-SC0021118. The work (O.M.F. and A.D.Y.) is supported by the Ministry of Science and Education of Ukraine (grant № РН/ 23 - 2023, "Influence of size effects on the electrophysical properties of graphene-ferroelectric nanostructures") at the expense of the external aid instrument of the European Union for the fulfillment of Ukraine's obligations in the Framework Program of the European Union for scientific research and innovation "Horizon 2020". E.A.E. and A.N.M also acknowledges the support of the National Academy of Sciences of Ukraine and CNMS (proposal ID: CNMS2024-A-02420).

## SUPPLEMENTARY MATERIALS

## Appendix A

Examples of how the LGD model works quantitatively are shown in **Figs. A1** and **A2** for polarization hysteresis loops in $Hf_xZr_{1-x}O_2$ thin films (abbreviated as HZO x/1-x) with a thickness of 9.2 nm for different contents of "x" Zr, from 100 % to 50 %. Fits of dielectric, paraelectric, antiferroelectric, and ferroelectric loops are shown. Blue symbols - hysteresis loops measured experimentally in the work of Park et al. [38]. Red solid curves − equilibrium polarization; purple dashed hysteresis loops − dynamic polarization calculated from the effective free energy of the LGD for the case $\lambda = 0$.

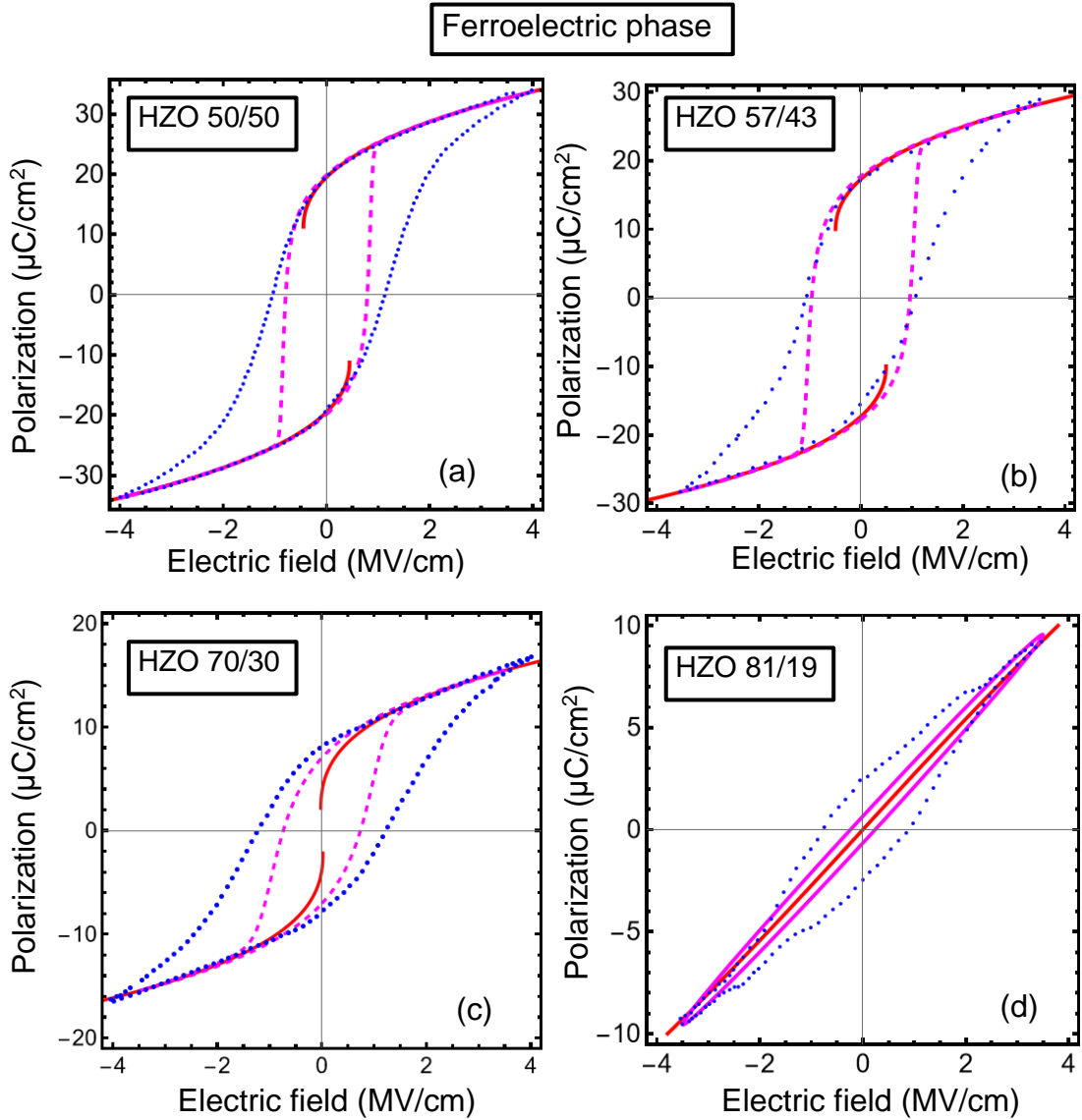

**Figure A1**. Polarization-field dependences (hysteresis loops) for the $Hf_xZr_{1-x}O_2$ films of thickness 9.2 nm (abbreviated as "HZO x/1-x") with the content of Zr 50 % **(a)**, 43 % **(b)**, 30 % **(c)** and 19 % **(d)**. Ferroelectric-type (a, b and c) and paraelectric-type (d) loops are shown. Red solid curves show the equilibrium polarization-

field dependences and magenta dotted curves show the dynamic polarization-field hysteresis loops calculated using the effective LGD model. Blue symbols represent the experimentally measured polarization-field dependences [38].

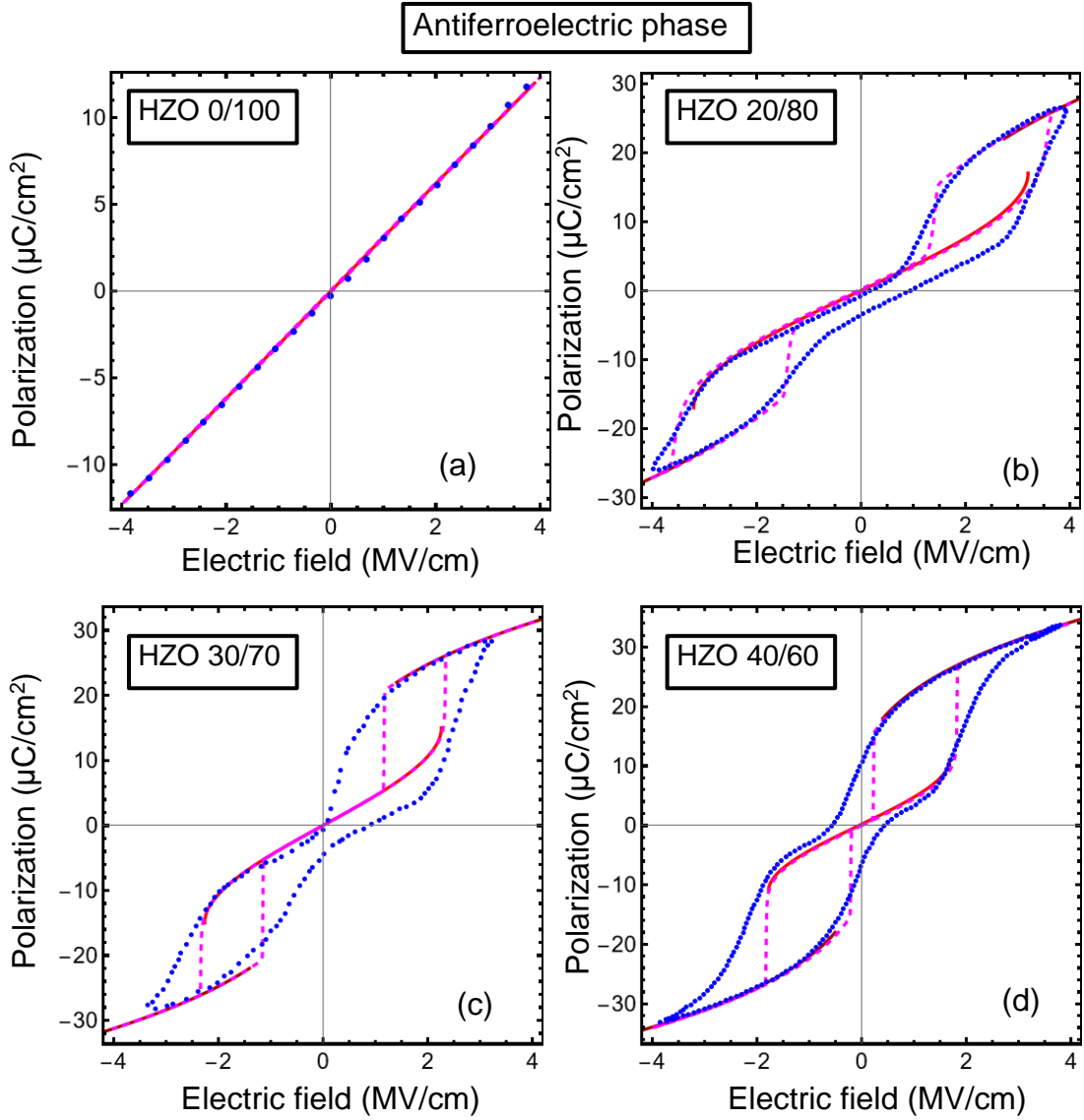

**Figure A2**. Polarization-field dependences (hysteresis loops) for the $Hf_xZr_{1-x}O_2$ films of thickness 9.2 nm (abbreviated as "HZO x/1-x") with the content of Zr 100 % **(a)**, 80 % **(b)**, 70 % **(c)** and 60 % **(d)**. Paraelectric-type (a) and antiferroelectric-type (b, c,d) loops are shown. Red solid curves show the equilibrium polarization-field dependences and magenta dotted curves show the dynamic polarization-field hysteresis loops calculated using the effective LGD model. Blue symbols represent the experimentally measured polarization-field dependences [38].

**Table AI.** Landau-Ginsburg-Devonshire parameters of $Hf_xZr_{1-x}O_2$ films with thickness $h = 9.2$ nm

| $x$ | phase | $P_r(x)$ $C/m^2$ | $P_{c1,2}(x)$ $C/m^2$ | $\tilde{\chi}(x)$ | $a_P(x)$ $\times 10^9$ m/F | $a_A(x)$ | $b_P \approx b_A$ $\times 10^{10}$ | $\eta(x)$ $\times 10^{10}$ |
|---|---|---|---|---|---|---|---|---|

| | | | | | ×10⁹ m/F | Vm⁵/C³ | Vm⁵/C³ |
|---|---|---|---|---|---|---|---|
| 0 | PE | 0 | N/A | 35 | 3.227 | N/P | N/P | N/P |
| 0.2 | AFE | 0 | 0.22, 0.17 | 40 | 0.812 | -0.936 | 0.9 | 1.93 |
| 0.3 | AFE | 0 | 0.22, 0.15 | 50 | 0.010 | -1.190 | 1.3 | 2.458 |
| 0.4 | AFE | 0 | 0.18, 0.11 | 48 | -0.104 | -0.937 | 1.1 | 2.888 |
| 0.50 | FE | 0.195 | N/A | 95 | -0.594 | N/P | 1.563 | N/P |
| 0.57 | FE | 0.173 | N/A | 76 | -0.743 | N/P | 2.483 | N/P |
| 0.70 | FE | 0.038 | N/A | 400 | -0.141 | N/P | 10.039 | N/P |
| 0.81 | PE | 0.025 | N/A | 31 | 3.643 | N/P | N/P | N/P |
| 1.00 | PE | 0 | N/A | 19 | 5.944 | N/P | N/P | N/P |

**Table A2.** The spontaneous order parameters, free energy densities, stability conditions of the thermodynamically stable spatially homogeneous phases of the free energy (1), and critical field(s). Adapted from Ref.[21]

| Phase and type of hysteresis loops | Spontaneous values of the order parameters | Free energy density and stability conditions | Coercive and/or critical field(s) $E_c$, and/or $P(E_c)$ * |
|---|---|---|---|
| Dielectric (DE) and paraelectric (PE) phases. Linear (DC) and paraelectric-type (PC) curves | $P_f = A_f = 0$ | $f_D = 0$ <br> $a_R > 0, a_A > 0$ | absent |
| Polar phase (FE) Ferroelectric-type single loops (SL) | $A_f = 0$, <br> $P_f = \pm\sqrt{-\frac{a_R}{b_P}}$, <br> $P_r \approx P_f$ | $f_P = -\frac{a_R^2}{4b_P}$ <br> $f_P = min, a_R < 0$, <br> $a_A b_P - \eta a_R > 0$ | $E_c = \pm\frac{2}{3\sqrt{3}}\frac{(-a_R)^{3/2}}{b_P}$ <br> $P_f(E_c) = 0$ |
| Mixed phase (FI) Pinched loops (PL) | $A_f = \pm\sqrt{-\frac{a_A b_P - \eta a_R}{b_A b_P - \eta^2}}$ <br> $P_f = \pm\sqrt{-\frac{a_R b_A - \eta a_A}{b_A b_P - \eta^2}}$ | $f_{PA} = \frac{-b_P a_A^2 - b_A a_R^2 + \eta a_P a_A}{4(b_A b_P - \eta^2)}$ <br> $f_{PA} = min$, <br> $a_A b_P - \eta a_R < 0, a_R b_A - \eta a_A < 0, b_A b_P > \eta^2$ | $E_c = \frac{\pm 2}{3\sqrt{3}}\frac{\left(-a_R+\frac{\eta a_A}{b_A}\right)^{3/2}}{b_P - \frac{\eta^2}{b_A}}$ <br> $P_f(E_c) = 0$ |
| Antipolar phase (AFE) Antiferroelectric-type double loops (DL) | $A_f = \pm\sqrt{-\frac{a_A}{b_A}}$, <br> $P_f = 0$. | $f_A = -\frac{a_A^2}{4b_A}$ <br> $f_A = min, a_A < 0$, <br> $a_R b_A - \eta a_A > 0$, <br> $\eta > \sqrt{b_A b_P}$ | $E_{c1} = \pm\left(a_R - \frac{a_A}{\eta}b_P\right)\sqrt{-\frac{a_A}{\eta}}$, <br> $P_{c1}(E_{c1}) = \pm\sqrt{-\frac{a_A}{\eta}}$ <br> $E_{c2} = \frac{\pm 2}{3\sqrt{3}}\frac{\left(a_R - \frac{\eta a_A}{b_A}\right)^{3/2}}{\frac{\eta^2}{b_A} - b_P}$ <br> $P_{c2}(E_{c2}) = \pm\sqrt{\frac{a_R b_A - \eta a_A}{3(\eta^2 - b_P b_A)}}$. |

* Noteworthy that $a_R(x, R) = a_P(x) + \frac{\varepsilon_0^{-1}}{\varepsilon_b + 2\varepsilon_s + \varepsilon_s(R/\lambda)}$.

Table A3. Rietveld refinement of crystallographic parameters of HfO$_{2-y}$ nanopowders

| Sample | Phase | Mass fraction, % | Scattering region sizes, nm | Lattice parameters | | | |
|---|---|---|---|---|---|---|---|
| | | | | $a$, Å | $b$, Å | $c$, Å | $\beta$, ° |
| 1 | m | 100 | 13 | 5,1229 | 5,1606 | 5,3005÷5,3007 | 99,1379÷99,1357 |
| 2 | m | 67,98 | 10 | 5,1408 | 5,1771 | 5,4113 | 100,0280 |
| | o61 | 32,02 | 9 | 9,9180 | 5,1452 | 5,1600 | - |
| 3 | m | 51,62 | 9 | 5,1248÷5,1251 | 5,1268÷5,1335 | 5,3101÷5,3079 | 98,8417÷98,8589 |
| | o61 | 48,38 | 14 | 10,0215÷10,0640 | 5,1339÷5,1280 | 5,0712÷5,0726 | - |
| 3 | m | 27,26 | 23 | 5,1520÷5,1515 | 5,2185÷5,2223 | 5,2813÷5,2804 | 98,9554÷98,9555 |
| | o61 | 72,74 | 10 | 10,1180÷10,1180 | 5,1793÷5,1834 | 5,1000÷5,1099 | - |